# Efficiency in managing peer-review of scientific manuscripts – editors' perspective

OLGICA NEDIĆ[1*], IVANA DRVENICA[2], MARCEL AUSLOOS[3,4] and ALEKSANDAR DEKANSKI[5]

[1]*Institute for the Application of Nuclear Energy - (INEP), University of Belgrade, Banatska 31b, Belgrade, Serbia (olgica@inep.co.rs)*
[2]*Institute for Medical Research, University of Belgrade, Dr Subotica 4, Belgrade, Serbia*
[3]*School of Business, University of Leicester, University Road, Leicester, United Kingdom*
[4]*Group of Researchers for Applications of Physics in Economy and Sociology (GRAPES), rue de la Belle Jardiniere 483, Angleur, Belgium*
[5]*Institute of Chemistry, Technology and Metallurgy, Department of Electrochemistry, University of Belgrade, Karnegijeva 4, Belgrade, Serbia*

***Abstract:*** The purpose of this paper is to introduce a model for measuring the efficiency in managing peer-review of scientific manuscripts by editors. The approach employed is based on the assumption that editorial aim is to manage publication with high efficiency, employing the least amount of editorial resources. Efficiency is defined in this research as a measure based on 7 variables. An on-line survey was constructed and editors of journals originating from Serbia regularly publishing articles in the field of chemistry were invited to participate. An evaluation of the model is given based on responses from 24 journals and 50 editors. With this investigation we aimed to contribute to our understanding of the peer-review process and, possibly, offer a tool to improve the "efficiency" in journal editing. The proposed protocol may be adapted by other journals in order to assess the managing potential of editors.

**RUNNING TITLE**: EFFICIENCY IN PEER-REVIEWING

*Corresponding author

## INTRODUCTION

The rate of scientific information generation has increased tremendously in the last few years. The authors generally perceive the speed of peer-review as slow.[1] The number of journals has also increased[2] and journal editors are facing an increasing number of submissions. The rate of increase in the number of researchers, studies and papers is far greater than the rate of increase in the number of journals, published pages or individuals involved in editorial activity. Due to this increased pace and the use of more informal approaches in workplace communication by modern technologies, as Smedley[3] explained, success is often determined by the individual management capacity. The main objective of editors is to publish good quality manuscripts that are free of errors. If this goal is achieved, the review process is effective. Editors are also expected to manage editorial work with high efficiency, *i.e.* employing the least amount of editorial resources. Effectiveness and efficiency do not necessarily correlate. Editors are required to be competent in dealing with authors, reviewers, associate editors, journal publishers and promotion, and also ethics.[4] Thus, in order to manage peer-review and publication process efficiently, journal editors define policies and develop strategies which include clearly stated aims and scope of a journal, guidelines for authors, ethical rules and guidelines for peer-reviewers, but they also apply implicit (personal) knowledge to develop a methodology to search for reviewers, evaluate reviewers' reports, and define criteria for making final decision.[5, 6] How to improve efficiency in scientific publishing has become a research field for journal editors.[7-9] Management of empirical, tacit, subjective knowledge seems to have the strongest impact on editorial strategy, even though one may propose objective technical helps.[10] Thus, the management performance of editors can be questioned from an "efficiency-defined" point of view. We propose to tackle this issue through a parsimonious model based on a finite size of editors in a specific domain, interrogated with the focus we just emphasize here above. With this investigation we aimed to contribute to our understanding of the peer-review process and, possibly, offer a tool for the evaluation and improvement of the efficiency in journal editing.

Thus, we propose a model to assess and measure efficiency in managing peer-review of scientific manuscripts. Although the term "efficiency" is either an economic or a thermodynamic concept and can be more firmly defined than it is done in this article, here it is defined through measures based on several appropriate variables. Efficiency in this research is understood as a measure to indicate the employment of editorial resources in order to manage submitted articles. Seven criteria are proposed (evaluated through multiple-choice questions) to define efficiency, as discussed below. The following aspects of the process were investigated: the number of invited reviewers, portion of invitations without response, portion of manuscripts for which a second round of reviewer invitation was needed, portion of inadequate reports (from the ethical point), portion of low quality reports (from the point of professional competence), timeliness of report submission and the way in which editors search for reviewers. Possible correlations between these variables were searched for, through a radar chart-like display from statistical means.

Peer-review management practices of editors of journals originating from Serbia regularly publishing articles in the field of chemistry and associated disciplines were analyzed through the proposed model. Twenty seven such journals were collected from the bibliographic databases (Web of Science Core Collection - WoS and the Serbian Citation Index - SCI). Some of them are managed by one person, whereas others have one editor-in-chief and subeditors (the initial information was found at journal websites). Since editors are positioned between authors who submit and external reviewers who evaluate manuscripts (although editors can also be reviewers), it seemed relevant to study and discuss the efficiency in handling scientific manuscripts in relation to the self-appreciated management skills of editors.

## EXPERIMENTAL

### *A model*

A model used to investigate efficiency in peer-review was based on 7 criteria: the number of invited reviewers, portion of invitations without response, portion of manuscripts for which a second round of reviewer invitation was needed, portion of inadequate reports, portion of low quality reports, timeliness of report submission and the way in which editors search for reviewers. Seven multiple-choice questions comprised a model questionnaire to assess this efficiency (Table I). For 6 questions, a choice of 4 responses is available, indicating different levels of efficiency (Table I, questions 1-6). The efficiency interval limits for each response were determined empirically. Since the absolute number of managed articles significantly varies between journals, in order to compare data between editors and journals, responses to questions are expressed as portions of the total number of processed articles. Thus, responses to 5 questions (1-4 and 6) relied on objective data, as they can be measured. The response to the question number 5 relies more on the subjective impression of the editor.

To each answer, defining a certain level of efficiency, a “weight factor” (WF) is assigned enabling the transformation of the data into simple numbers which can be further statistically analyzed. For example, for the question “How many reviewers do you invite in the first round?”, if the answer is >4, this response is assigned WF = 1 (WF1) indicating the lowest level of efficiency; if the answer is 1 or 2, this response is assigned WF = 4 (WF4) indicating the highest level of efficiency. In general, WF = 4 recognizes the most efficient occurrence, indicating the least employment of editorial resources in order to manage submitted articles (such as the number of invitations to reviewers and the actual number of responses, the number of adequate reports, and the time needed to obtain them).

Table I. Seven multiple-choice questions used to estimate peer-review efficiency; for 6 questions only one quantitative response could be chosen; each response was assigned a weight factor (WF); for the 7th question more than one qualitative response could be chosen and there were no WFs

| Question | Weight factor | | | |
|---|---|---|---|---|
| | WF1 | WF2 | WF3 | WF4 |
| How many reviewers do you invite in the first round? | >4 | 4 | 3 | 1-2 |
| What is the portion of manuscripts for which a second round of reviewer invitation is needed? | >60% | 41-60% | 25-40% | <25% |
| What is the portion of invitations to reviewers without response? | >60% | 41-60% | 25-40% | <25% |
| What is the portion of inadequate reports? | >10% | 6-9% | 3-5% | 1-2% |
| How do you estimate the quality of reports? | Predominantly poor | Equivalent number of good and poor | Predominantly good | Good |
| How do you estimate the timeliness of report submission? | >10 days after deadline | <10 days after deadline | On time | Before deadline |

| 7. How do you search for reviewers? |
|---|
| I invite a colleague who was already a reviewer for this journal |
| I invite a colleague who was an author of article in this journal |
| I invite a colleague whom I know personally |
| I use bibliographic databases (WoS, SCOPUS, Google Scholar, PubMed) |
| I review manuscripts frequently by myself |
| Other (please, state how) |

In the last question (Table I, question 7), the method used for finding reviewers is taken into consideration. More than one qualitative answer can be chosen; there are no WFs for it. Finally, in order to evaluate how the duration of the editorial activity influences the peer-review efficiency, a question on how long the person has been an editor is introduced.

An on-line survey was constructed with two parts: (I) and (II); editors were invited to participate by e-mail. After the first call, a reminder was sent 2 weeks later to those who did not respond and, again, 2 weeks later. After the third call, no more answer was requested nor received. The survey started on November 1st 2015, and lasted for 6 weeks (general information on journals was collected in October 2015).

(I) In the first part of the questionnaire, the surveyed editors were asked to identify their editorial role. Editors-in-chief were further directed to general questions on the journal they manage, such as number of subeditors, number of members in the editorial board, journal position in WoS or SCI database, number of printed articles per year, language of publication and mode of financing.

(II) In the second part of the survey, all participants were asked about their personal practice and outcomes on several aspects which contribute to peer-review efficiency and depend on management skills of editors.

*Calculation of peer-review efficiency*

After assigning WFs to the answers (Table I, questions 1-6), an “overall efficiency” of the peer-review process managed by an editor can be calculated from his/her responses. The overall efficiency ($E$) of peer-review activity managed by one editor (or in one journal) is estimated in two ways. The first one results from the calculation of an arithmetic mean value (average WF) for the 6 WFs (i.e. responses to 6 questions) characteristic for a particular journal. The second efficiency measure takes into consideration the area (expressed in arbitrary units, AU) of the hexagon (drawn as a radar chart) constructed for each journal using its 6 individual WFs. The choice of axes for the hexagon construction follows the order of questions (from 1 to 6).

The overall efficiency ($E$) for each editor is, finally, expressed as the percentage of the maximal efficiency ($E_{max}$). Two $E_{max}$ values are calculated, one for each approach for data presentation: $E_{1max}$ corresponding to the maximal arithmetic mean WF = 4 (i.e. all 6 individual WFs are 4), while $E_{2max}$ is reached when the relative area of a hexagon is maximal (i.e. defined by 6 WFs which are all equal to 4). Thus, two $E$ values are calculated for each editor: $E_1$ from the mean WF and $E_2$ from the hexagon area.

Notice that when a journal is managed by only one editor, the calculated efficiency corresponds to the peer-review efficiency of this particular journal. When several editors are responsible for the peer-review process in the same journal, WFs are determined for each editor and then average WFs are calculated as mean values for that set of editors in order to obtain average WFs for the particular journal. Although this data processing reduces the accuracy to some extent, it is necessary to enable a comparison between different journals.

*Study population to test a model*

The number of chemical and chemistry-associated journals included in this study was 27. The list was made by using bibliographic databases: the Web of Science Core Collection (WoS) for extracting international journals referenced in InCites Journal Citation Reports (11 of them) and the Serbian Citation Index (SCI) for identifying journals referenced only in national citation index (16 of them). A scope and contents of more than 100 journals were investigated and the final list was made after checking topics of published articles (especially in journals without words “chemistry” or “chemical” in their titles). Information on each journal was initially searched for on its website (in October 2015), collecting the name(s) of editor(s) (in-chief and subeditors). The most important filtering criterion for inclusion in the list was that journals are regularly published (over several years, including 2015). The list of journals is given in Table II (titles of some national journals are translated into English). The number of journals having only editor-in-chief is 12, whereas 15 journals are managed by an editor-in-chief and subeditor(s). Seventeen journals publish articles only in English, 7 both in English and Serbian and 3 only in Serbian. The editorial population involved in the study is 70.

Table II. List of journals involved in the study, their referencing in InCites Journal Citation Reports (JCR) and the number of responses received through the survey

| Name of the journal (ISSN / eISSN)) | Referencing in JCR<br>Category (Rank/Number of journals) | No Responses/<br>Invitations |
|---|---|---|
| Nuclear Technology and Radiation Protection (1451-3994 / 1452-8185) | Nuclear Science & Technology (25/34) | 1/1 |
| Thermal Science (0354-9836 / 334-7163) | Thermodynamics (25/55) | 2/4 |
| Chemical Industry and Chemical Engineering Quarterly (1451-9372 / 2217-7434) | Chemistry, Applied (48/72)<br>Engineering, Chemical (89/135) | 1/4 |
| Hemijska industrija (Chemical Industry)(0367-598X / 2217-7426 -) | Engineering, Chemical (121/135) | 6/9 |
| International Journal of Electrochemical Science (- / 1452-3981) | Electrochemistry (21/28) | 1/1 |
| Journal of Medical Biochemistry (1452-8258 / -) | Biochemistry & Molecular Biology (257/290) | 1/2 |
| Journal of Mining and Metallurgy, Section B: Metallurgy (1450-5339 / 2217-7175) | Metallurgy & Metallurgical Engineering (35/74) | 1/2 |
| Journal of the Serbian Chemical Society 0352-5139 / 1820-7421 | Chemistry, Multidisciplinary (114/157) | 14/16 |
| Kragujevac Journal of Science (1450-9636 / 2466-5509) | Uncategorized | 1/1 |
| MATCH Communications in Mathematical and in Computer Chemistry (0340-6253 / -) | Chemistry, Multidisciplinary (80/157)<br>Computer Sci. Interdisc. Appl. (45/102)<br>Mathematics, Interdisc. Appl. (28/99) | 2/2 |
| Science of Sintering (0350-820X / 1820-7413) | Materials Science, Ceramics (14/26)<br>Metallurgy & Metallurgical Eng. (49/74) | 1/2 |
| Vojnotehnički glasnik (Military Technical Journal) (0042-8469 / 2217-4753) | | 1/1 |
| Facta Universitatis - Series: Physics, Chemistry and Technology (0354-4656 / -) | | 2/2 |
| Hemijski pregled (Chemical Overview) (0440-68267 / -) | | 1/1 |
| Acta Periodica Technologica (1450-7188 / 2406-095X) | | 1/2 |
| Arhiv za farmaciju (Archive for Pharmacy) (0004-1963 / 2217-8767) | | 1/2 |
| Bakar (Copper) (0351-0212 / - ) | | -/1 |
| Metallurgical and Materials Engineering (2217-8961 / - ) | | -/1 |
| Processing and Application of Ceramics (1820-6131 / 2406-1034) | | 1/1 |
| Reciklaža i održivi razvoj (Recycling and Sustainable Develoment) (1820-7480 / 2560-3132) | | 2/3 |
| Savremene tehnologije (Advanced Technologies) (2217-9712 / -) | | 1/1 |
| Scientific Technical Review (1820 0206 / -) | | 1/2 |
| Svet polimera (World of Polymers) (1450-6734 / -) | | -/1 |
| Tehnika (Technics) (0040-2176 / 2560-3086) | | 3/3 |
| Voda i sanitarna tehnika (Water and Sanitary Technics) (0350-5049 / -) | | 1/1 |
| Zaštita materijala (Material Protection) (0351-9465 / 2466-2585) | | 3/3 |
| Zbornik Matice srpske za prirodne nauke (Matica Srpska Journal of Natural Sciences) (0352-4906 / -) | | 1/1 |

*Data analysis*

Collected data were analyzed for individual editors (journals), together for all participants or subdivided into groups: editors in WoS and SCI journals. One very specific journal *(Journal of the Serbian Chemical Society*, *JSCS*) was also selected for a “horizontal” comparison of editorial practices and outcomes between subeditors, as 14 responses were received from its editors. Statistical analysis was performed by using SPSS software to check normality of the data distribution. Correlations between different components of peer-review efficiency were searched for (a correlation was assumed to be strong when the Pearson’s correlation coefficient, $r$ was $\geq 0.75$). Statistically significant differences (at $P < 0.05$) between groups of editors: in (a) WoS journals, (b) SCI journals and (c) subeditors in the *JSCS* were assessed by using the Mann-Whitney U test. In order to test the coherence of the two measures, $E_1$ and $E_2$, between groups, the entire set of data (for all editors or journals) was additionally analyzed along a rank-size law methodology and the Kendall $\tau$ rank correlation measure.

## RESULTS

*Response rate*

Out of 70 invited editors, 50 responded; 22 editors-in-chief and 28 subeditors (30 males and 20 females). A response rate of 71.4 % is considered satisfactory for the social non-mandatory surveys.[11,12] Out of 27 surveyed journals, information was collected for 24, *i.e.* 88.9 %.

*Calculation of peer-review efficiency in WoS journals*

Weight factors related to particular responses from WoS journal editors are shown in Table III (journals are presented as letters, as formal permission to identify editors or journals with specific results was not obtained, except for the *JSCS*). In 6 WoS journals, there were only editors-in-chief; all of them responded. In 5 WoS journals, there were subeditors beside the editor-in-chief; not all of them responded. In order to compare data between WoS journals, WFs for journals having several editors were averaged at a journal level by calculating mean values from answers provided by individual (sub)editors.

As explained in the Experimental section, the overall efficiency of the peer-review activity in one journal was estimated in two ways. The first one resulted from the calculation of the arithmetic mean value (average WF) for 6 WFs corresponding to responses characteristic for a particular editor or a journal. The second efficiency measure relied on the area of the hexagon drawn by using 6 individual WFs as axis for each journal (Table III and Figure 1). Both efficiency measures were further expressed as percentages of the maximal efficiency: $E_1$ for the mean value and $E_2$ for the hexagon area.

Table III. Efficiency of peer-review process estimated by (sub)editors in WoS journals

| | Weight Factor | | | | | | | | | | |
|---|---|---|---|---|---|---|---|---|---|---|---|
| **Journal** | **Number of reviewers invited in the first round** | **Portion of manuscripts for which a second round of reviewer invitation is needed** | **Portion of invitations to reviewers without response** | **Portion of inadequate reports** | **Quality (competence) of reports** | **Timeliness of report submission** | **Sum** | **Average** | **$E_1$ / %** | **Relative surface area of hexagon (AU)** | **$E_2$ / %** |
| **A** | 4 | 3.4 | 3.8 | 3 | 3.4 | 2.4 | 20.0 | 3.33 | **83.25** | 28.5 | **68.51** |
| **B** | 3.1 | 3.1 | 3.2 | 1.9 | 3.1 | 2.6 | 17.0 | 2.83 | **70.75** | 22.5 | **54.09** |
| **C** | 2 | 1.5 | 2.5 | 2.5 | 2.5 | 3 | 14.0 | 2.33 | **58.25** | 14.2 | **34.13** |
| **D** | 4 | 4 | 4 | 3 | 2 | 2 | 19.0 | 3.17 | **79.25** | 26.1 | **62.74** |
| **E** | 3 | 3 | 3 | 4 | 3 | 1 | 17.0 | 2.83 | **70.75** | 20.8 | **50.00** |
| **F** | 4 | 4 | 3 | 4 | 4 | 3 | 22.0 | 3.67 | **91.75** | 34.6 | **83.17** |
| **G** | 4 | 3 | 4 | 4 | 3 | 3 | 21.0 | 3.50 | **87.50** | 31.6 | **75.96** |
| **H** | 2 | 1 | 1 | 4 | 3 | 4 | 15.0 | 2.50 | **62.50** | 16.9 | **40.62** |
| **I** | 4 | 4 | 4 | 2 | 3 | 2 | 19.0 | 3.17 | **79.25** | 26.0 | **62.50** |
| **J** | 4 | 4 | 4 | 4 | 4 | 3 | 23.0 | 3.83 | **95.75** | 38.1 | **91.59** |
| **K** | 4 | 4 | 4 | 2 | 3 | 3 | 20.0 | 3.33 | **83.25** | 29.0 | **69.71** |
| **Mean** | 3.46 | 3.18 | 3.32 | 3.13 | 3.09 | 2.64 | 18.82 | 3.135 | **78.386** | 26.21 | **63.00** |
| **SD** | 0.815 | 1.051 | 0.939 | 0.910 | 0.577 | 0.779 | 2.822 | 0.4710 | **11.775** | 7.265 | **17.466** |
| **CV** | 0.236 | 0.331 | 0.283 | 0.291 | 0.187 | 0.295 | | | **0.150** | | **0.277** |
| **Max** | 4 | 4 | 4 | 4 | 4 | 4 | 24.0 | 4.00 | **100.00** | 41.57 | **100.00** |

By analyzing Table III and especially Figure 1, it becomes obvious that the efficiency of peer-review, from the moment of reviewer invitation to the moment of report collection, is differentially affected by the examined components of the process in each journal. This finding suggests the existence of a major personal influence of an editor on the final outcome. Although efficiencies $E_1$ and $E_2$ are highly correlated, as expected, the method used to define $E_2$ is more illustrative for the comparison of peer-review efficiencies between editors or journals. A diagrammatic presentation of data by radar charts offers a better overview of the strengths and weaknesses of a process in a particular journal or managed by particular editor than numbers read from a table. For example, it can be seen from Table III that $E_1$ is the same in journals A and K, yet WF values which define $E_2$ differ significantly (see Table III and Figure 1). Editors in journal A are the least efficient in obtaining review reports on time, while the efficiency in journal K is mostly affected by the judgment of an editor that there are too many inadequate reports. Thus, by using a model proposed in this article, editors/journals can obtain an insight in specific weaknesses which need better management. In general, editors are the least satisfied with the timeliness of review reports and this variable decreases the overall efficiency in the majority of journals.

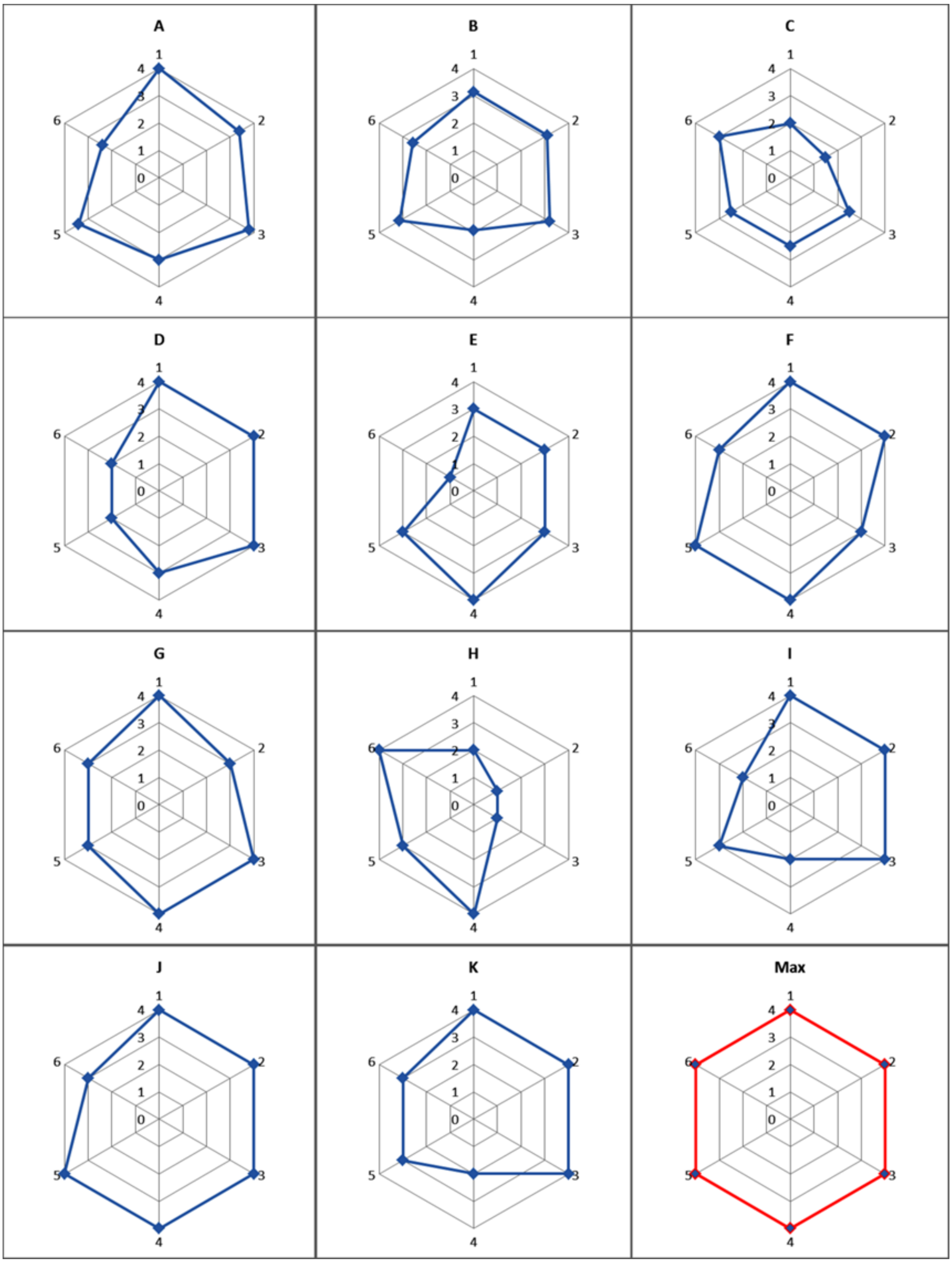

Figure 1. Efficiency ($E_2$) of the peer-review process in WoS journals, estimated via hexagon construction, using a 6 weight factor scheme for each journal (A-K).

*Data analysis*

A statistical analysis was performed to find correlations between the investigated parameters, and between the $E_1$ and/or $E_2$ values within the WoS group of editors/journals. Only a few strong correlations were found (with a Pearson's correlation coefficient, $r \geq 0.75$). Positive correlations were found between: (a) the number of reviewers invited in the first round and the portion of manuscripts for which a second round of reviewer invitation was needed ($r = 0.93$), (b) the number of reviewers invited in the first round and the portion of invitations without response ($r = 0.88$), (c) the number of reviewers invited in the first round and the average WF ($r = 0.84$), and (d) $E_1$ and $E_2$ ($r = 0.98$).

In the second part of the data analysis, the $E$ values were correlated with the years spent as being an editor and the number of approaches applied to search for reviewers (Figure 2a). The results for journals having more than one (sub)editor were, again, averaged to allow some global comparison - although we are aware that mean values are a compromise, not exact data. The following was found: (a) no correlation is seen between the efficiency and the duration of editorial experience (no editor was less than 7 years on duty) and (b) employing more ways to search for reviewers contributes to the efficiency ($r = 0.75$, Figure 2a). No correlation emerges between a particular way(s) used to search for reviewers and the peer-review efficiency in WoS journals.

*Comparison of peer-review efficiency between WoS, SCI journals and one journal managed by several subeditors*

Peer-review efficiency was investigated in the same manner as described above in another two sets of samples: editors in SCI journals and subeditors in the *JSCS* (Table IV, Figure 2b and 2c, and Supplement 1). Similar relations profiled from the data on the number of reviewers invited in the first round and the portion of invitations without response for SCI journals ($r = 0.86$) as for WoS journals. The correlation between efficiency and the number of ways used to find reviewers in SCI journals, however, was much weaker than in WoS journals ($r = 0.48$). When responses from 14 subeditors in *JSCS* were analyzed, only one strong correlation emerged: between the number of reviewers invited in the first round and the portion of invitations without response ($r = 0.76$). In contrast to the first two groups of editors, in this last case, a weak negative correlation was detected between the efficiency and the number of ways used to find reviewers ($r = -0.42$). The number of years having been in editorial activity or the particular reviewer invitation pattern was not directly related to the efficiency in either group.

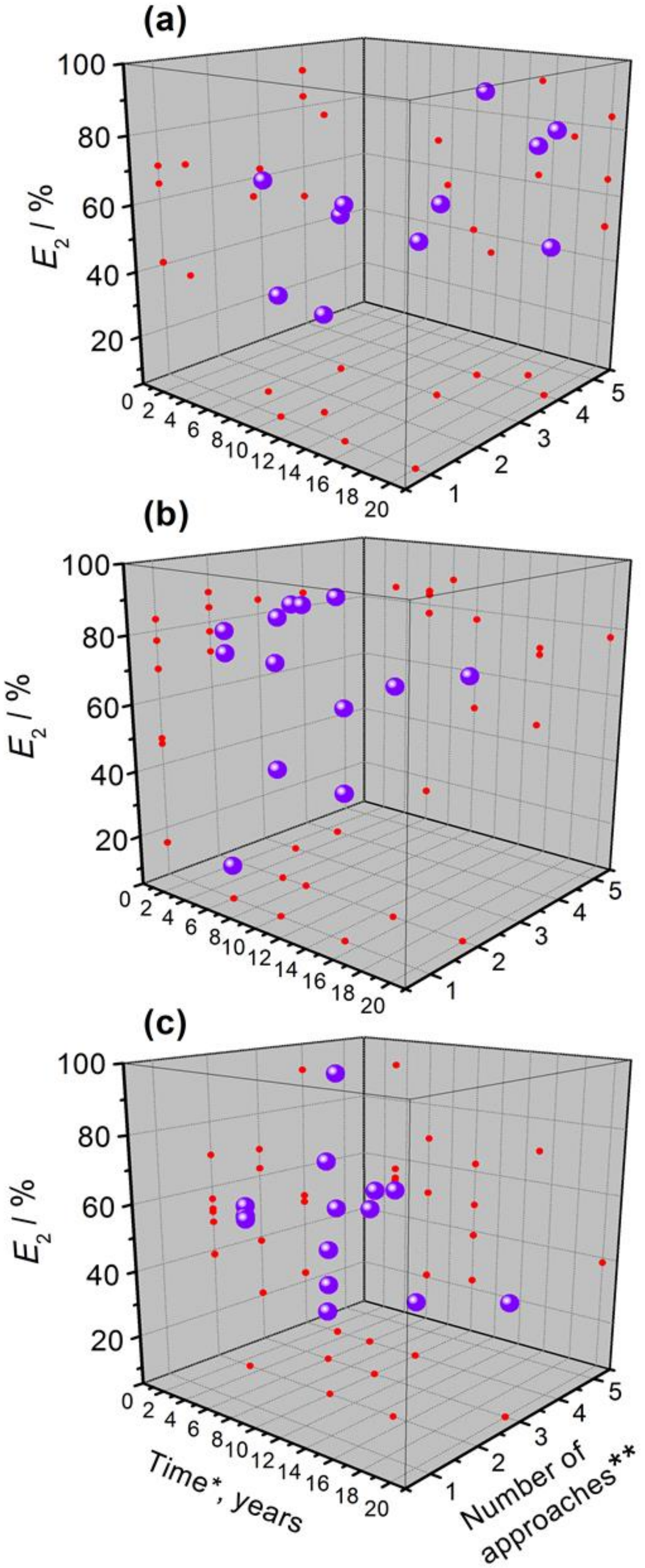

Figure 2. Relation between peer-review efficiency ($E_2$), the number of years having been in editorial activity (*) and the number of approaches applied to search for reviewers (**) in: (a) WoS journals, (b) SCI journals and (c) within one WoS journal, *i.e. JSCS*. Big purple dot represents a combined result for one (sub)editor taking into consideration efficiency ($E_2$) estimated by that (sub)editor (vertical axis), number of years he/she is being in editorial activity (left horizontal axis) and number of approaches he/she applies to search for reviewers (right horizontal axis). Small (red) dots represent 3-dimensional projections of big (purple) dots.

The Mann-Whitney U test was used to assess possible difference between the results ($E_1$ and $E_2$) obtained for three groups of data. No statistically significant difference in the efficiency was seen between editors in WoS and SCI journals, or between all editors in WoS journals and subeditors in the *JSCS*. There was, however, a significant difference in the efficiency between SCI journals and subeditors in the *JSCS*. Editors in national journals, in general, scored higher than subeditors in the *JSCS*.

Table IV. Efficiency ($E_1$) and ($E_2$) of the peer-review process estimated by editors in SCI journals (left hand side) and by subeditors in one WoS journal (*JSCS*, right hand side)

| SCI journals | | | | | | Journal of the Serbian Chemical Society | | | | | |
|---|---|---|---|---|---|---|---|---|---|---|---|
| Jour-nal | Sum | Average WF | $E_1$ / % | RSA* (AU) | $E_2$ / % | Sub-editor | Sum | Average WF | $E_1$ / % | RSA* (AU) | $E_2$ / % |
| A’ | 9.5 | 1.58 | 39.58 | 6.2 | 15.10 | A’’ | 17.0 | 2.83 | 70.75 | 20.7 | 49.76 |
| B’ | 21.0 | 3.50 | 87.50 | 32.0 | 77.08 | B’’ | 12.0 | 2.00 | 50.00 | 8.7 | 20.91 |
| C’ | 22.0 | 3.67 | 91.67 | 35.0 | 84.38 | C’’ | 23.0 | 3.83 | 95.75 | 38.1 | 91.59 |
| D’ | 23.0 | 3.83 | 95.83 | 38.1 | 91.67 | D’’ | 18.0 | 3.00 | 75.00 | 22.1 | 53.12 |
| E’ | 17.0 | 2.83 | 70.83 | 19.9 | 47.92 | E’’ | 17.0 | 2.83 | 70.75 | 21.6 | 51.92 |
| F’ | 22.0 | 3.67 | 91.67 | 34.6 | 83.33 | F’’ | 20.0 | 3.33 | 83.25 | 29.0 | 69.71 |
| G’ | 21.0 | 3.50 | 87.50 | 32.0 | 77.08 | G’’ | 15.0 | 2.50 | 62.50 | 16.0 | 38.46 |
| H’ | 20.0 | 3.33 | 83.33 | 28.5 | 68.75 | H’’ | 15.0 | 2.50 | 62.50 | 16.0 | 38.46 |
| I’ | 17.0 | 2.83 | 70.83 | 19.2 | 46.30 | I’’ | 19.0 | 3.17 | 79.25 | 26.0 | 62.50 |
| J’ | 20.0 | 3.33 | 83.33 | 29.4 | 70.83 | J’’ | 17.0 | 2.83 | 70.75 | 19.9 | 47.84 |
| K’ | 21.0 | 3.50 | 87.50 | 32.0 | 77.08 | K’’ | 20.0 | 3.33 | 83.25 | 28.6 | 68.75 |
| L’ | 22.7 | 3.78 | 94.44 | 36.9 | 88.89 | L’’ | 17.0 | 2.83 | 70.75 | 20.4 | 49.04 |
| M’ | 22.0 | 3.67 | 91.67 | 35.0 | 84.38 | M’’ | 12.0 | 2.00 | 50.00 | 9.5 | 22.84 |
| | | | | | | N’’ | 18.0 | 3.00 | 75.00 | 23.4 | 56.25 |
| Mean | 19.862 | 3.309 | 82.745 | 29.14 | 70.215 | Mean | 17.14 | 2.856 | 71.393 | 21.43 | 51.511 |
| SD | 3.643 | 0.6085 | 15.173 | 9.011 | 21.666 | SD | 3.009 | 0.5007 | 12.519 | 7.742 | 18.610 |
| CV | | | 0.1836 | | 0.3086 | CV | | | 0.1753 | | 0.3613 |
| Max | 24.0 | 4.00 | 100.00 | 41.6 | 100.00 | | 24.0 | 4.00 | 100.00 | 41.57 | 100.00 |

RSA* - Relative surface area of hexagon expressed in arbitrary units (AU)

*The coherence of the efficiency measures*

In order to further test the coherence of the two measures, $E_1$ and $E_2$, the entire set of data (for all editors or journals) was additionally analyzed along a rank-size law methodology and a Kendall $\tau$ rank correlation measure. The results of the former analysis are shown in Figure 3. Other figures can be displayed. To save space, and to make our point, we only propose these two figures: one for the efficiency $E_1$ and the other for $E_2$, with different types of “best fits”, a power or a linear law; other figures can be easily imagined from these. The figures illustrate much regularity, far from the usual power law expectation, and closer to a straight line best fit: this is due to the fact that the number of data points only spans a decade (of editors or journals). Nevertheless, some fine agreement is observed. This is confirmed, in some sense, by the Kendall $\tau$ rank correlation measure which is respectively equal to 0.972, 0.973 and 0.949. Even though these values look close to each other, one can observe that the variation between editors in WoS and SCI journals is quite weak, but the measure about subeditors somewhat differs.

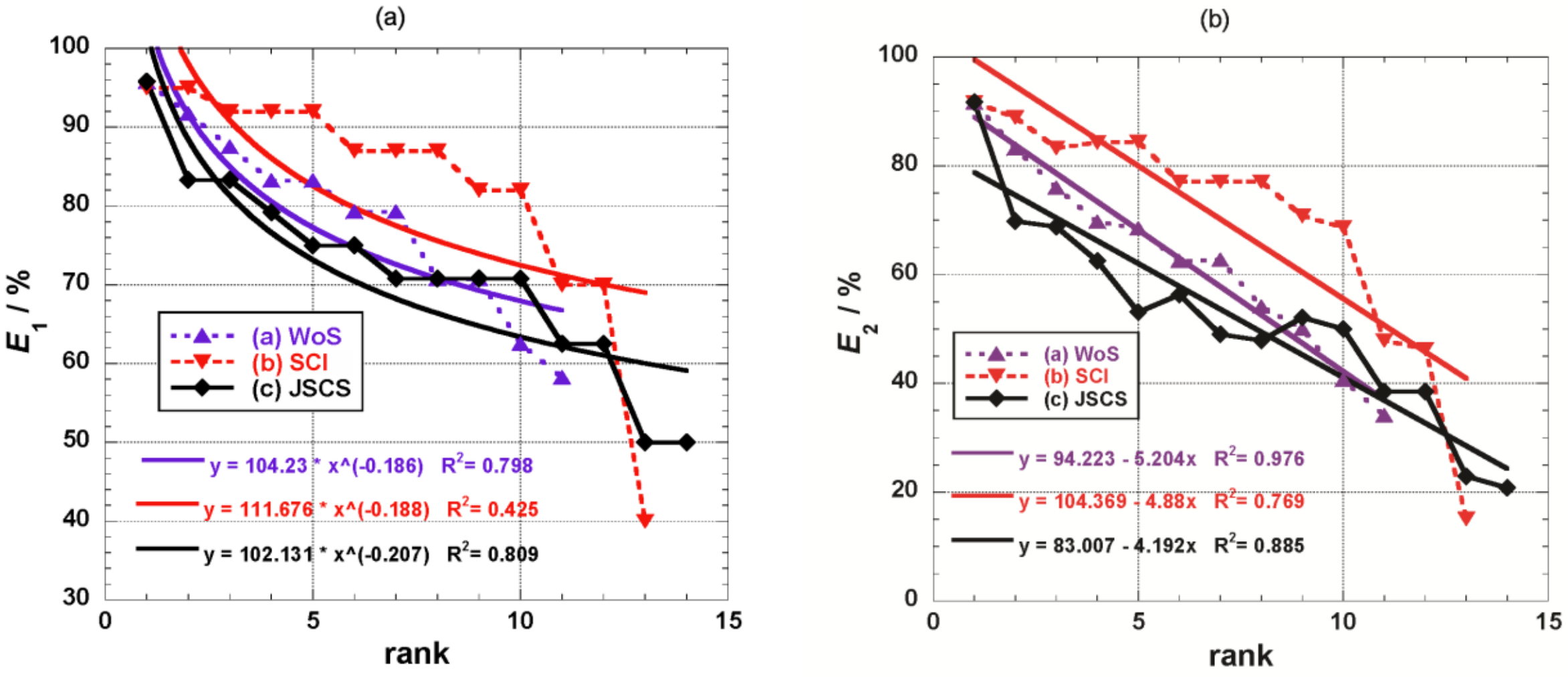

Figure 3. Rank-size law for the efficiency ($E_1$) and ($E_2$), with power law and linear fits, respectively, for editors in: (a) WoS journals, (b) SCI journals and (c) JSCS, distinguished by symbols: triangles on base, triangles on tip and diamonds. The best respective (the least mean square procedure) fits are given.

*An overview of the data for individual editors*

Finally, in Figure 4 an overview of the data for several measures characteristic for each individual (sub)editor is given: (a) calculated efficiency of peer-review process, (b) number of years spent in editorial activity and (c) personal approach in searching for reviewers. Editors are grouped by alphabetical order of their names, not by journal affiliation (there was no particular pattern when analyzed by journal affiliation). As it can be seen, most editors (37/50 editors) use databases to search for reviewers; approximately half of them invite colleagues whom they know (23/50) or who already reviewed for their journals (22/50); several editors invite previous authors to become reviewers (15/50); several editors review manuscripts by themselves (11/50); whereas few editors (8/50) employ other strategies for peer-review (such as a panel of reviewers, an invitation of a reviewer recommended by an editor’s colleague or a reviewer suggested by an author). By examining the data in Figure 4, it becomes obvious that there is no specific invitation pattern (*i.e.* specific combination of approaches in searching for reviewers) which results in more efficient peer-review process, as assessed from the chosen efficiency measures.

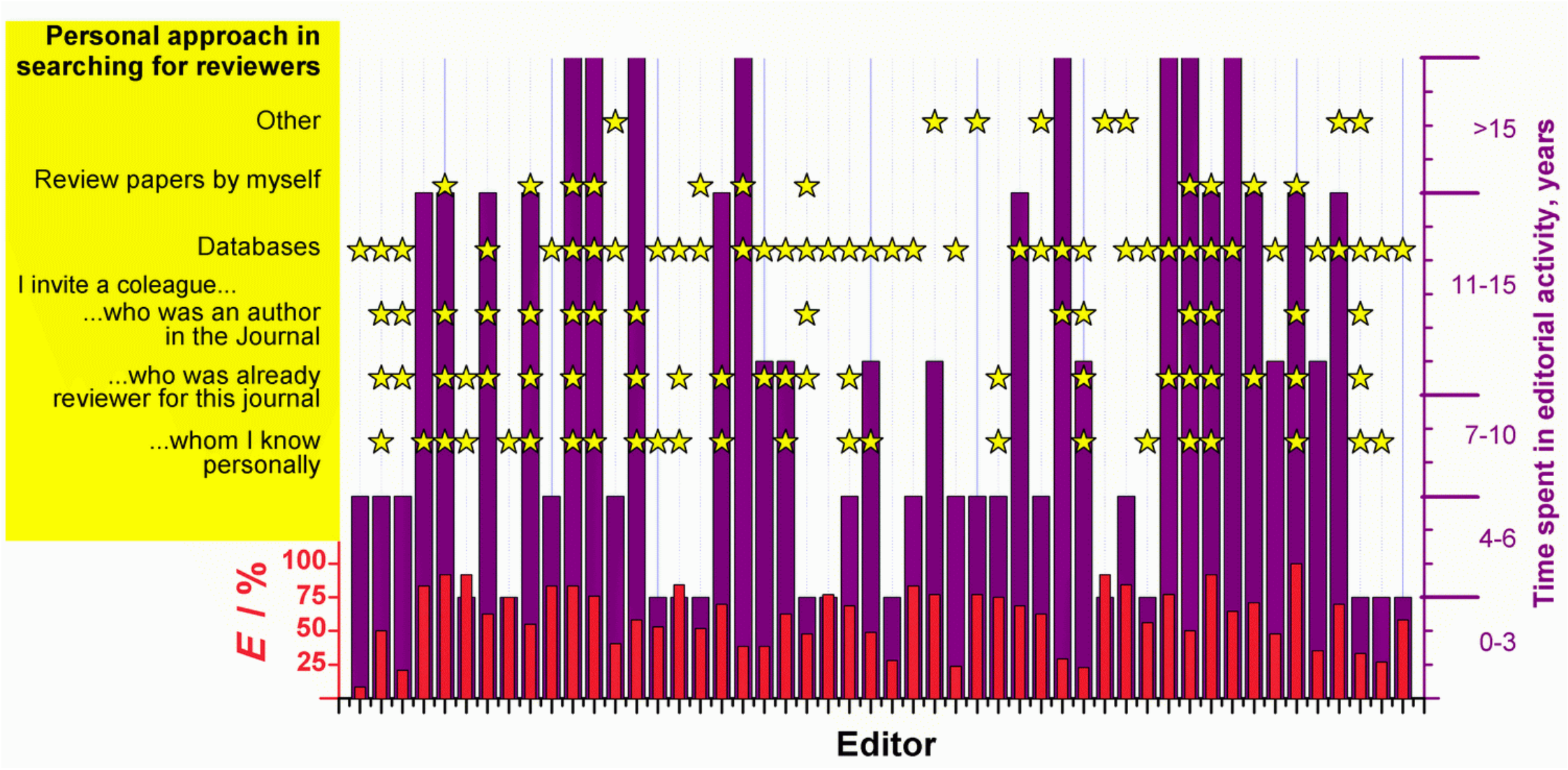

Figure 4. The relation between peer-review efficiency $E_2$ (pink column), number of years spent in editorial activity (purple column) and personal approach in searching for reviewers for individual editors (yellow stars positioned at 6 levels correspond to 6 approaches in searching for reviewers listed on the left hand side).

## DISCUSSION

Before becoming editors, most researchers spent years being authors and reviewers of scientific papers, gathering personal experience and knowledge on this subject. Once they become editors, they are expected to use their knowledge to manage editorial work and up-grade it in order to achieve high quality and efficiency in publishing papers mainly by others. In other words, knowledge lifecycle in scientific journals is very similar to the one seen in traditional business process, it can be also related to supply chain management,[13, 14] and is relying very often on just one or few people. As Del-Ray-Chamorro and colleagues[15] proposed, knowledge management domain presents an added value to other management techniques. Measurement of knowledge management performance is a serious challenge, and there are only few published articles on this topic.[15-18] According to Yu and colleagues[18], one of the main reasons for the lack of such studies is the unavailability of effective and quantitative methods for measuring values generated from the knowledge management system. We hope to contribute to this issue somewhat.

In this paper, a model for measuring efficiency of peer-review in scientific journals is indeed introduced. Application of the proposed model to assess efficiency managed by journal editors confirmed that the present methodology of editors can be questioned from a "practical efficiency-defined" point of view and an outcome can be evaluated and measured after transformation of the survey data into simple numerical indicators. Although each single outcome can be specifically analyzed in relation to a single editor and tied to specific advantages and weaknesses which that editor exhibits while managing editorial work, more general implications have been discussed in our report.

From the above results, it can be seen that a similar degree of overall (un)efficiency was recorded in WoS and SCI journals, with no statistically significant difference between them. Data on the number of subeditors in one journal, however, did not completely resemble data on editors in different journals. This variation, when looking at journals on one hand, and subeditors in the *JSCS*, on the other hand, is confirmed when reading the mean values in Table III and Table IV. It may be conjectured that this finding is indicating that although there is a general editorial policy in one journal, subeditors manage editorial activity mostly in an individual manner and in accordance with personal experience and knowledge. By examining Figure 2, one can see that WoS journals are managed mostly by experienced editors (being at least 7 years at this position), whereas SCI journals are managed by greater number of editors who have spent fewer years on duty. Although the difference in peer-review efficiency between two groups of editors was not statistically significant, editors in SCI journals scored slightly higher level of efficiency. As it was previously stated that editors are the least satisfied with the timeliness of reports submission, this factor can significantly influence efficiency. In this model, however, duration of the expected peer-review period defined by editors was not taken into consideration. It might be accounted for in further studies.

Statistically significant difference in the efficiency measures $E_1$ and $E_2$ was seen between *JSCS* subeditors and editors in SCI journals, but not between *JSCS* subeditors and group of editors in WoS journals. Of course, one must bear in mind that responses from subeditors in the *JSCS* are included in the dataset of responses from all editors in WoS journals, which contributes to some extent to greater agreement of results. By comparing two efficiency measures $E_1$ and $E_2$, it seems that $E_2$-approach, relying on hexagon presentation of data, is more helpful in estimating peer-review efficiency in a journal. For example, peer-review processes identified as the most efficient overall (with the highest $E$ values, such as in journals F and G, Figure 1) differed in the efficiency of separate components, as can be clearly seen from the radar charts.

Common to all three editorial groups (WoS, SCI and subeditors in the *JSCS*) is a positive correlation between the number of reviewers invited in the first round and the portion of invitations to reviewers without response. This finding can be explained by a frequent invitation of “reliable” or “known” reviewers, who tend to accept the invitation and send a report. In 4 out of 11 WoS journals (D, I, J and K, Table III), for example, editors responded that they ask 1-2 reviewers to review manuscripts, whereas the portion of manuscripts for which a second round of reviewer invitation was needed is less than 25 %. The portion of invitations without response in these journals is also less than 25 %. Editors of these journals seem to have developed some efficient strategy to find reviewers. It was previously recognized that when editors invite well-known people to review, they expect a high-quality report[5]. Additionally, an internationally well-known editor may know many more potential reviewers, and because of reputation and network, finding good and active reviewers may be much easier than in the case of less well-known editors.

On the other hand, in 2 WoS journals (C and H, Table III), the overall efficiency is rather low; the greatest problems editors are facing are related to the necessity to initially invite 4 reviewers on average, who often do not respond, leading to a significant number of second round invitations. In general, indeed, it is hard to find scientists in specific research fields with sufficient expertise in peer-review and who have time to review.[19] Rejection to review is sometimes justified – if potential reviewers feel that they are not competent enough or do not have enough time for professional review, if they personally know or are related to author(s) and if they have some conflicts of interest.[20-22] Editors in the journal *Annals of Emergency Medicine* have developed a specific stratifying system to divide their pool of reviewers into 3 categories according to their scoring on the number of reviews performed, their timeliness of report submission and quality.[23] At the end, 55 % of invitations were sent to top-class reviewers, which contributed to the pool of reviewers by 25 %. The most important outcome was a significant decrease in late reports, thus, an obvious improvement in the "efficiency".

Other variables which were not taken into consideration in this study and in the proposed model, but may influence the response rate of reviewers in a particular journal, are the number of submitted manuscripts, the number of manuscripts which are peer-reviewed (not desk-rejected) and the number of reviews performed by an individual reviewer. It may be expected that in journals with fewer submissions it is less hard to complete a peer-review process "efficiently". The reputation of a journal and the acknowledgment for reviewing (from a journal or a publisher) can also contribute to the response rate. As already stated in Introduction section, effectiveness and efficiency do not always correlate. For example, some journals will regularly invite at least three referees, which in our model makes them less efficient, but ensures greater effectiveness in selecting high quality articles.

Most answers in our survey relied on objective "parameters", which could be measured and quantified by numbers, whereas the one on the quality (competence) of reports was based on a subjective impression. Editors were asked to judge on the quality of submitted reports, although for the purpose of this survey, we did not define what is considered to be a "good review report", either through quantitative[24] or qualitative measures[25, 26], thereby allowing each editor to personally appreciate and measure what "quality" means. Thus, we are aware that answers expressing some greater dissatisfaction could have also reflected more stringent criteria on the quality exerted by certain editors. Another point should be also highlighted, although it was not an intended subject of the imagined model - editorial behavior. Wang et al.[27] have found that in the case of biased editors, the effect on the quality of peer-review process is even worse than the effect of biased reviewers. In the same spirit, one might consider the effect of coercive citations in peer-review process efficiency.[28]

## CONCLUSIONS

Our objective in this paper was to propose a parsimonious model relying on 7 elements in order to measure peer-review efficiency in scientific journals. The model was tested through a rather large set of editors, though necessarily of the limited size and in a specific field, but it is expected to be of wider application. Even though other variables can contribute to efficiency, the proposed protocol may be adapted by other journals in order to assess managing potential of editors. A similar degree of overall (un)efficiency was recorded in WoS and SCI journals. In general, editors are the least satisfied with the timeliness of review reports. A positive correlation between the number of reviewers invited in the first round and the portion of invitations without response was found, suggesting frequent invitation of "reliable" or "known" reviewers, who accept the invitation and send a report. No correlation was seen between the efficiency and the duration of editorial experience. Employing more ways to search for reviewers, however, contributes to the efficiency.

*Acknowledgments*: This paper is part of scientific activities in COST Action TD1306 New Frontiers of Peer Review (PEERE).

## ИЗВОД

## Ефикасност у рецензирању научних радова – из угла уредника

ОЛГИЦА НЕДИЋ[1], ИВАНА ДРВЕНИЦА[2], MARCEL AUSLOOS[3,4] и АЛЕКСАНДАР ДЕКАНСКИ[5]

[1]*Институт за примену нуклеарне енергије (ИНЕП), Универзитет у Београду, Србија,* [2]*Институт за медицинска истраживања, Универзитет у Београду, Србија*, [3]*School of*

*Business, University of Leicester, Велика Британија,* [4]*Group of Researchers for Applications of Physics in Economy and Sociology (GRAPES), Angleur, Белгија* и [5]*Институт за хемију, технологију и металургију, Одељење за електрохемију, Универзитет у Београду, Србија*

У овом раду је описан модел за мерење ефикасности у рецензирању научних радова, који могу применити уредници. Приступ теми је подразумевао да је циљ уредника да управља процесом публиковања што ефикасније, уз што мању употребу ресурса. Ефикасност је дефинисана коришћењем 7 променљивих. Креирана је електронска анкета и уредници часописа који редовно излазе у Србији, а чија је тема хемија и сродне дисциплине, су позвани да је попуне. Модел је евалуиран на основу одговора 50 уредника из 24 часописа. Предлагањем овог модела, желели смо да допринесемо разумевању процеса рецензирања и да понудимо „алат“ за евентуално побољшање ефикасности у уређивачкој делатности. Предложени протокол могу усвојити часописи у циљу утврђивања управљачких способности уредника.

REFERENCES

1. V. M. Nguyen, N. R. Haddaway, L. F. G. Gutowsky, A. D. M. Wilson, A. J. Gallagher, M. R. Donaldson, N. Hammerschlag, S. J. Cooke, *PloS One* **10** (2015) e0132557 (https://doi.org/10.1371/journal.pone.0139783)

2. X. Gu, K. L. Blackmore KL, *Scientometrics* **108** (2016) 693 (https://doi.org/10.1007/s11192-016-1985-3)

3. J. Smedley J, *OR Insight* **22** (2009) 221 (https://doi.org/10.1057/ori.2009.11)

4. J. Galipeau, V. Barbour, P. Baskin, S. Bell-Syer, K. Cobey, M. Cumpston, J. Deeks, P. Garner, H. MacLehose, L. Shamseer, S. Straus, P. Tugwell, E. Wager, M. Winker, D. Moher, *BMC Medicine* **14** (2016) 16 (https://doi.org/10.1186/s12916-016-0561-2)

5. E. Roohi, O. Mahian, *Sci. Engin. Ethics* **21** (2015) 809 (https://doi.org/10.1007/s11948-014-9549-5)

6. J. A. Garcia, R. Rodriguez-Sanchez, J. Fdez-Valdivia, *Scientometrics* **113** (2017) 45 (https://doi.org/10.1007/s11192-017-2470-3)

7. T. Jefferson, M. Rudin, S. B. Folse, F. Davidoff, *Cochrane Database Syst. Rev.* **2** (2007) MR000016 (https://doi.org/10.1002/14651858.MR000016.pub3)

8. K. Anderson, *Inform. Serv. Use* **35** (2015) 171 (https://doi.org/10.3233/ISU-150776)

9. A. Weiskittel, *Math. Comp. Forest. Nat.-Res. Sci.* **7** (2015) 81 (http://mcfns.net/index.php/Journal/article/view/MCFNS7.2_4)

10. M. J. Mrowinski, A. Fronczak, P. Fronczak, O. Nedic, M. Ausloos, *Scientometrics* **107** (2016) 271 (https://doi.org/10.1007/s11192-016-1871-z)

11. K. B. Sheehan, *J. Comp.-Mediat. Commun.* **6** (2001) JCMC621 (https://doi.org/10.1111/j.1083-6101.2001.tb00117.x)

12. SurveyMonkey, http://s3.amazonaws.com/SurveyMonkeyFiles/Response_Rates.pdf (2009) accessed 23 Nov. 2015

13. A. Gunasekaran, C. Patel, E. Tirtiroglu E, *Int. J. Operat. Product. Manag.* **21** (2001) 71 (https://doi.org/10.1108/01443570110358468)

14. P. Charan, R. Shankar, R. K. Baisya, *Business Process Manag. J.* **14** (2008) 512 (https://doi.org/10.1108/14637150810888055)

15. F. M. Del-Rey-Chamorro, R. Roy, B. van Wegen, A. Steele, *J. Knowl. Manag.* **7** (2003) 46 (https://doi.org/10.1108/13673270310477289)

16 J. Swaak J, A. Lansink, E. Heeren, B. Hendriks, P. Kalff, J-W. den Oudsten, R. Bohmer, R. Bakker, C. Verwijs, 59th AEPF-Tagung conference, Bremen, Germany, 2000

17. W. D. Yu, P. L. Chang, S. J. Liu, International Symposium on Automation and Robotics in Construction 2006 (ISARC 2006), Tokyo, Japan, Proceedings, 2006, p. 124 (https://doi.org/10.22260/ISARC2006/0026)

18. W. D. Yu, P. L. Chang, S. H. Yao, S. J. Liu, *Construct. Manag. Econom.* **27** (2009) 733 (https://doi.org/10.1080/01446190903074978)

19. R. K. F. Clark, *Br. Dent. J.* **213** (2012) 153 (https://doi.org/10.1038/sj.bdj.2012.721)

20. L. Tite, S. Schroter, *J. Epidemiol. Commun. Health* **61** (2007) 9 (https://doi.org/10.1136/jech.2006.049817)

21. C. D. Bailey, D. R. Hermanson, T. J. Louwers, *J. Account. Edu.* **26** (2008) 55 (https://www.jstor.org/stable/41948838)

22. M. A. Zaharie, C. L. Osoian, *Eur. Manag. J.* **34** (2016) 69 (https://doi.org/10.1016/j.emj.2015.12.004)

23. S. M. Green, M. L. Callaham, *Ann. Emerg. Med.* **57** (2011) 149 (https://doi.org/10.1016/j.annemergmed.2010.08.005)

24. M. Ausloos, O. Nedic, A. Fronczak, P. Fronczak, *Scientometrics* **106** (2016) 347 (https://doi.org/10.1007/s11192-015-1705-4)

25. C. A. Geithner, A. N. Pollastro, *Adv. Physiol. Educ.* **40** (2016) 38 (https://doi.org/10.1152/advan.00071.2015)

26. M. Lamont, J. Guetzkow, *Quality Is Recognized by Peer Review Panels: The Case of the Humanities*, in *Research Assessment in the Humanities*. M. Ochsner, S. E. Hug, H-D. Daniel, Eds., Springer, Berlin, Germany, 2016, p. 31 (https://doi.org/10.1007/978-3-319-29016-4_4)

27. W. Wang, X. Kong, J. Zhang, Z. Chen, F. Xia, X. Wang, *SpringerPlus* **5:903** (2016)

(https://doi.org/10.1186/s40064-016-2601-y)
28. C. Herteliu, M. Ausloos, B. V. Ileanu, G. Rotundo, T. Andrei, *Publications* 5 (2017) 15 (https://doi.org/10.3390/publications5020015)

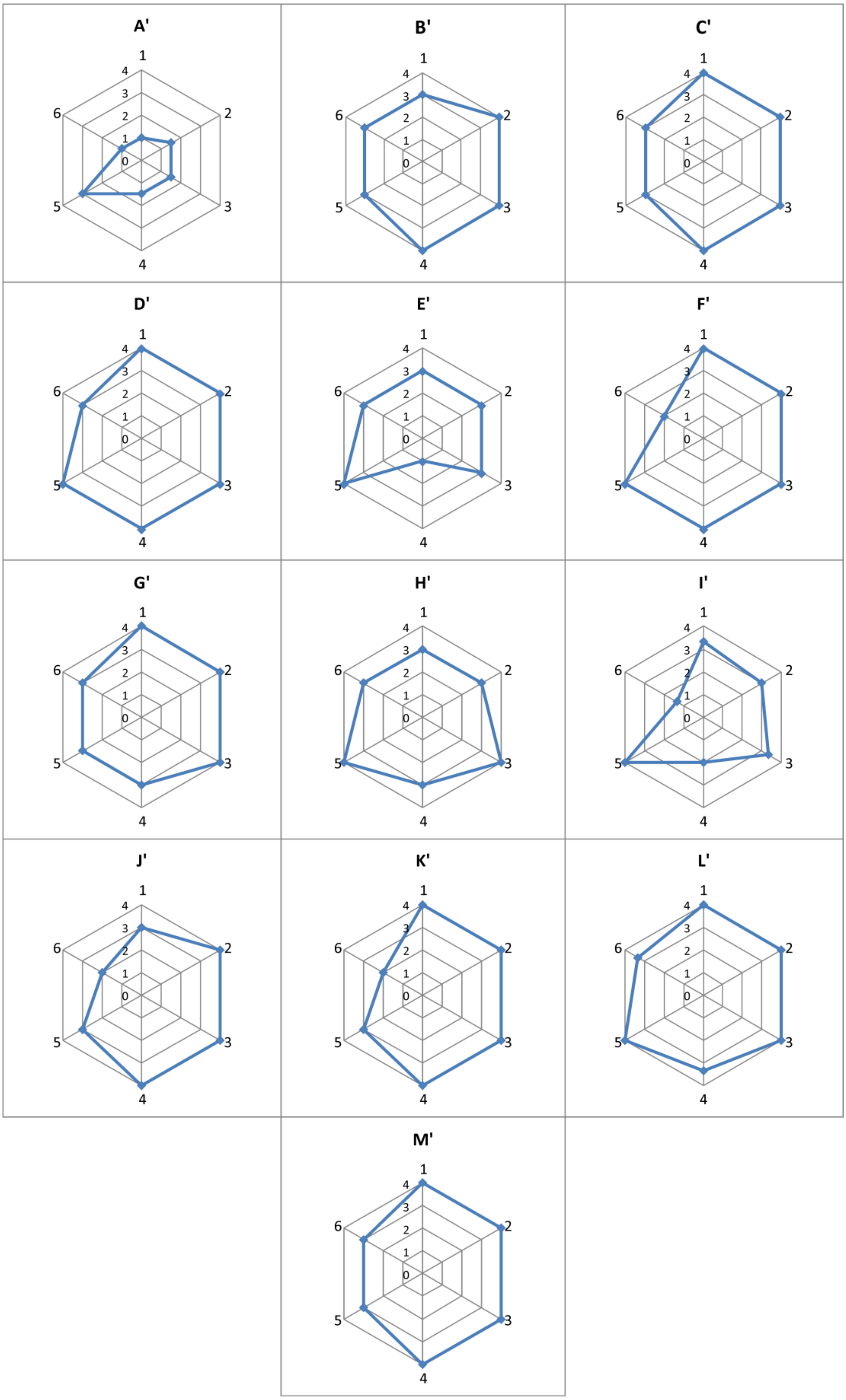

Supplement 1. Efficiency ($E_2$) of the peer-review process in SCI journals, estimated via hexagon construction, using a 6 weight factor scheme for each journal (A'-M').